\documentclass[12pt,a4paper,final]{iopart}
\usepackage{iopams, bm}
\usepackage[perpage]{footmisc}
\usepackage{graphicx}
\usepackage{color}
\usepackage[inline]{enumitem}
\usepackage{multirow}
\usepackage{siunitx}

\begin{document}

\title{New techniques for a measurement of the electron's electric dipole moment}

\author{C~J~Ho, J~A~Devlin\footnote{Present address: CERN, Esplanade des Particules 1, CH-1211 Geneva 23, Switzerland}, I~M~Rabey\footnote{Present address: Max-Planck-Institut f{\"u}r Quantenoptik, Hans-Kopfermann-Stra{\ss}e 1, D-85748 Garching, Germany}, P~Yzombard\footnote{Present address: Kastler Brossel Laboratory, Sorbonne Universit\'{e}, CNRS, ENS Universit\'{e} PSL, Coll\`{e}ge de France, 4 place Jussieu, case 74, 75252, Paris cedex 05, France}, J~Lim, S~C~Wright, N~J~Fitch, E~A~Hinds, M~R~Tarbutt, B~E~Sauer}

\address{Centre for Cold Matter, Blackett Laboratory, Imperial College London, Prince Consort Road, London SW7 2AZ UK}

\eads{christopher.ho15@imperial.ac.uk, ben.sauer@imperial.ac.uk}

\begin{abstract}
The electric dipole moment of the electron (eEDM) can be measured with high precision using heavy polar molecules. In this paper, we report on a series of new techniques that have improved the statistical sensitivity of the YbF eEDM experiment. We increase the number of molecules participating in the experiment by an order of magnitude using a carefully designed optical pumping scheme. We also increase the detection efficiency of these molecules by another order of magnitude using an optical cycling scheme. In addition, we show how to destabilise dark states and reduce backgrounds that otherwise limit the efficiency of these techniques. Together, these improvements allow us to demonstrate a statistical sensitivity of $\SI{1.8e-28}{e.cm}$ after one day of measurement, which is 1.2 times the shot-noise limit. The techniques presented here are applicable to other high-precision measurements using molecules.
\end{abstract}

\vspace{2pc}

\section{Introduction}

Any local, energy-positive, Lorentz-invariant field theory must conserve CPT, the combined symmetry of parity (P), time-reversal symmetry (T) and charge conjugation (C)~\cite{Lehnert2016, Greenberg2006}. A permanent electric dipole moment of the electron (eEDM) violates both P and T, and so requires some amount of CP violation to exist. The Standard Model already includes a small amount of CP violation, leading to a very small eEDM: $d_e \sim \SI{e-44}{e.cm}$~\cite{Pospelov2014}. However, more CP violation is required to explain the observed asymmetry between matter and antimatter in the universe~\cite{Sakharov1991,Cline2007}. Theoretical models such as supersymmetry attempt to explain beyond-Standard-Model physics by introducing new sources of CP violation. These lead to larger values for the eEDM: $d_e \sim \SIrange[range-phrase = -]{e-27}{e-30}{e.cm}$~\cite{Pendlebury2000}, which are accessible to modern precision measurement techniques. Therefore, searching for the eEDM can restrict these new theories, and can also help reveal new physics beyond the Standard Model.

Experiments that search for the eEDM measure the energy shift due to the linear Stark interaction of the eEDM with an electric field. Such experiments typically use paramagnetic atoms or molecules which greatly enhance the strength of this interaction. These systems are also sensitive to any electron-nucleon interaction that violates P and T, but here we adopt the single-source approach of assuming that any P, T-violating signal is due to the eEDM. The energy shift for a paramagnetic system can be expressed as $\Delta E = -d_e E_\text{eff}$, where $E_\text{eff}$ is an effective electric field calculated from atomic and molecular theory which can be much larger than applied fields in the laboratory. For atoms, it turns out that $| E_\text{eff} | \approx 8 Z^3 \alpha^2 E_\text{ext}$, where $Z$ is the nuclear charge, $\alpha$ is the fine structure constant and $E_\text{ext}$ is the externally applied electric field~\cite{Hinds1997}. The dependence of the interaction energy on $Z^3$ implies that heavy atoms are much better at measuring $d_e$, and indeed early measurements used heavy atoms such as Cs and Tl~\cite{Murthy1989,Regan2002}, which have enhancement factors of $E_\text{eff}/E_\text{ext} = 120$ and $-585$ respectively\footnote{The sign indicates whether the atomic EDM is parallel ($+$) or antiparallel ($-$) to the eEDM.}. The linear dependence of $E_\text{eff}$ on $E_\text{ext}$ indicates that the atoms are only weakly polarised in the external electric field. In polar molecules, the interaction energy is larger because it is easier to polarise these molecules in an electric field. It is more appropriate to write $E_\text{eff} = \eta E_\text{eff,max}$ for molecules, where $\eta$ is the degree of polarisation of the molecule, and $E_\text{eff,max}$ is the maximum effective field seen by the electron when the molecule is fully polarised, $\eta = 1$. The latter is typically in the range \SIrange{10}{100}{GV \per cm}, which is much larger than electric fields that can be applied in the laboratory.

In 2011, the precision of atomic measurements was surpassed in an experiment using YbF, setting a new upper limit\footnote{All eEDM limits quoted here are at the 90\% confidence level.} of $| d_e | < \SI{1.06e-27}{e.cm}$~\cite{Hudson2011}. The enhancement of YbF was $E_\text{eff} \approx \SI{-14.5}{GV \per cm}$. Crucially, a systematic effect which is large for atoms -- the Zeeman interaction with the motional magnetic field mimicking the EDM interaction -- is highly suppressed in molecules due to their strong tensor polarisability~\cite{Hudson2002}. In 2014, the ACME collaboration pushed the limit down to $| d_e | < \SI{8.7e-29}{e.cm}$ using a beam of ThO molecules in an $\Omega$-doublet state~\cite{Baron2013}. Molecules in this state are fully polarised in a small applied electric field, giving $E_\text{eff} = E_\text{eff,max} \approx \SI{84}{GV \per cm}$. The $\Omega$-doublet can also be used conveniently for internal co-magnetometry. In 2017, a measurement using trapped HfF$^+$ molecular ions ($E_\text{eff} \approx \SI{23}{GV \per cm}$) reported the limit $| d_e | < \SI{1.3e-28}{e.cm}$~\cite{Cairncross2017}. This experiment benefited from the long coherence times available in a molecular ion trap, but was limited by the relatively low number of ions trapped. In 2018, the ACME collaboration improved on their limit, reaching $| d_e | < \SI{1.1e-29}{e.cm}$~\cite{Andreev2018}. This last result constrains any new physics arising from T-violating effects to energy scales above \SI{3}{TeV}~\cite{Andreev2018}.

Many new ideas are now emerging on how to improve measurement sensitivity even further. In present experiments, the total uncertainty is typically dominated by the statistical uncertainty, so methods to improve statistical sensitivity are needed. Laser cooling is a technique that can significantly extend coherence times and brightness of atomic or molecular beams, thereby lowering the statistical sensitivity. The use of laser cooling is evident in proposed eEDM experiements using atoms such as Fr~\cite{Inoue2014}, diatomic molecules such as YbF~\cite{Tarbutt2013} or BaF~\cite{Aggarwal2018}, or polyatomics such as YbOH~\cite{Kozyryev2017}. Transverse laser cooling of molecular beams of YbF and YbOH has been demonstrated~\cite{Lim2018, Augenbraun2020} and experiments with ultracold molecules in beams, fountains, or traps now seem feasible. Other ideas include the use of polar molecules embedded in a rare-gas matrix~\cite{Vutha2018}. In this paper, we demonstrate new state preparation and state detection techniques using $^{174}$YbF molecules, which increase the statistical sensitivity to the eEDM by a factor of 20 over our previous methods described in~\cite{Kara2012}. The new techniques implemented here can also be applied to other experiments using molecules and, since they rely strongly on scattering many photons from a particular molecular state, are also closely related to the techniques used to apply laser cooling to molecules~\cite{Tarbutt2018}.

\section{Measurement method}
\label{sec:method}

\begin{figure}[tb]
    \centering
    \includegraphics{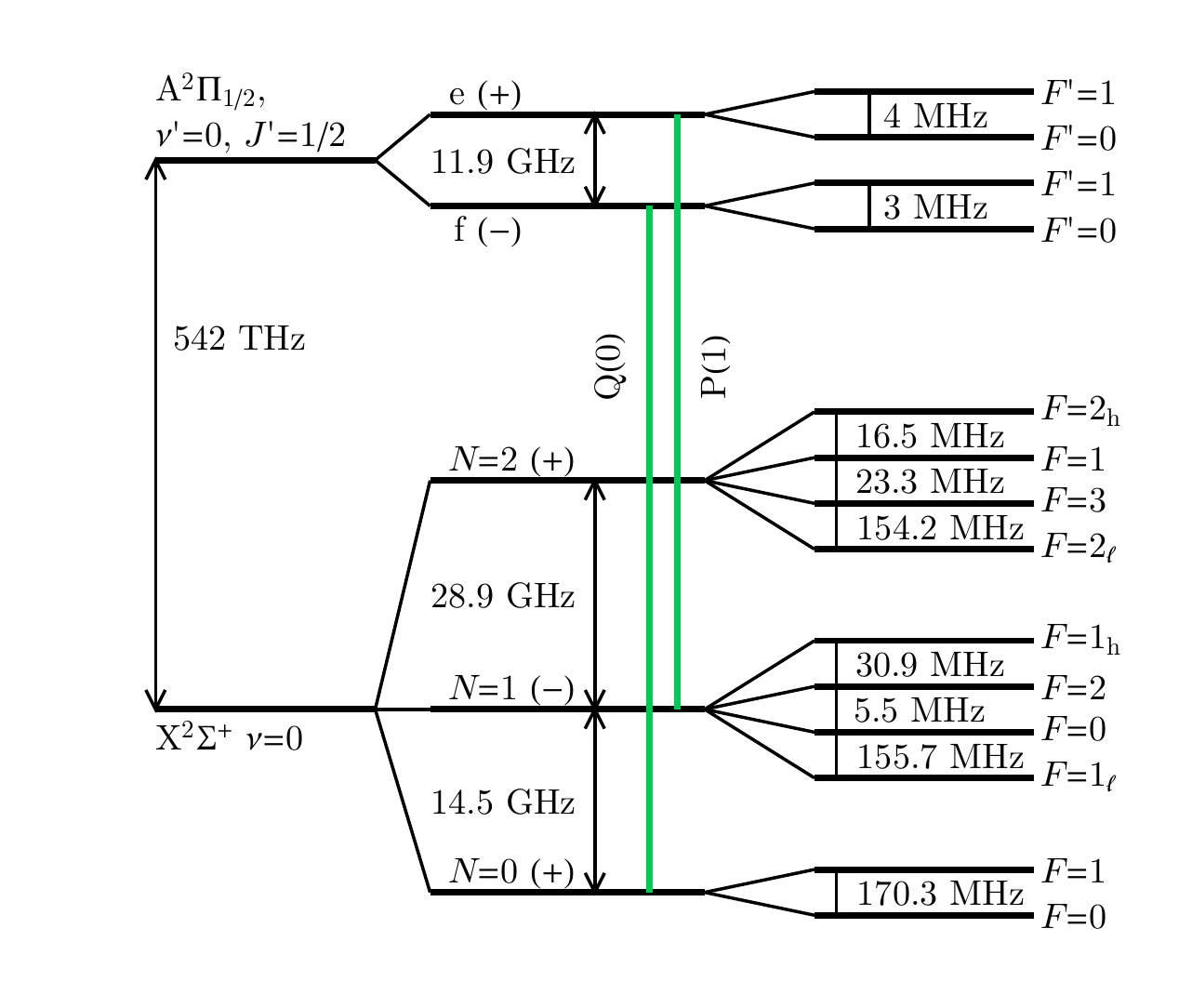}
    \caption{Relevant states of $^{174}$YbF and their energy separations (not to scale). The parity of the states is indicated in parentheses. Q(0) and P(1) refer to optical transitions as described in the text.}
    \label{fig:energyLevelDiagram}
\end{figure}

The relevant energy levels of $^{174}$YbF for the new state preparation and detection schemes are shown in Fig.~\ref{fig:energyLevelDiagram}. We use the molecular population in the lowest three rotational states of the ground electronic and vibrational state, $\text{X}^2\Sigma^+ (\nu = 0, N = 0, 1, 2)$, where $\nu$ and $N$ are vibrational and rotational quantum numbers, respectively. The parity of the rotational states is given by $(-1)^N$, and is indicated in parentheses in Fig.~\ref{fig:energyLevelDiagram}. The interaction between the electron spin, the spin-1/2 fluorine nucleus, and the molecule rotation splits the rotational states into hyperfine states with total angular momenta $F = N+1, N, N, N-1$. Because there are two hyperfine states with $F=N$, we distinguish them using the notation $N_\text{h}$ and $N_\ell$ to refer to the states that lie higher and lower in energy, respectively. For $N=0$, the molecule rotation is absent and so the hyperfine states are simply $F=0, 1$. Each hyperfine level has ($2F+1$) Zeeman sublevels. The $N=0$ state is used for the eEDM measurement, as will be discussed later. In the experiment, we excite molecules to the electronically excited state $\text{A}^2\Pi_{1/2} (\nu' = 0, J'=1/2)$, where $J'$ is the total electronic angular momentum of the excited state. This state is split into an $\Omega-$doublet, labelled by $e$ and $f$, which have parities $+1$ and $-1$~\cite{Sauer1999}. These are further split by the hyperfine interaction into states $F' = 0, 1$, but these are typically unresolved. To ease notation, we use Q(0) and P(1) to refer to optical transitions connecting $N=0$ to $f$ and $N=1$ to $e$ respectively, as shown in Fig.~\ref{fig:energyLevelDiagram}.

\begin{figure}[p]
    \vspace{9.64mm}
    \centering
    \includegraphics{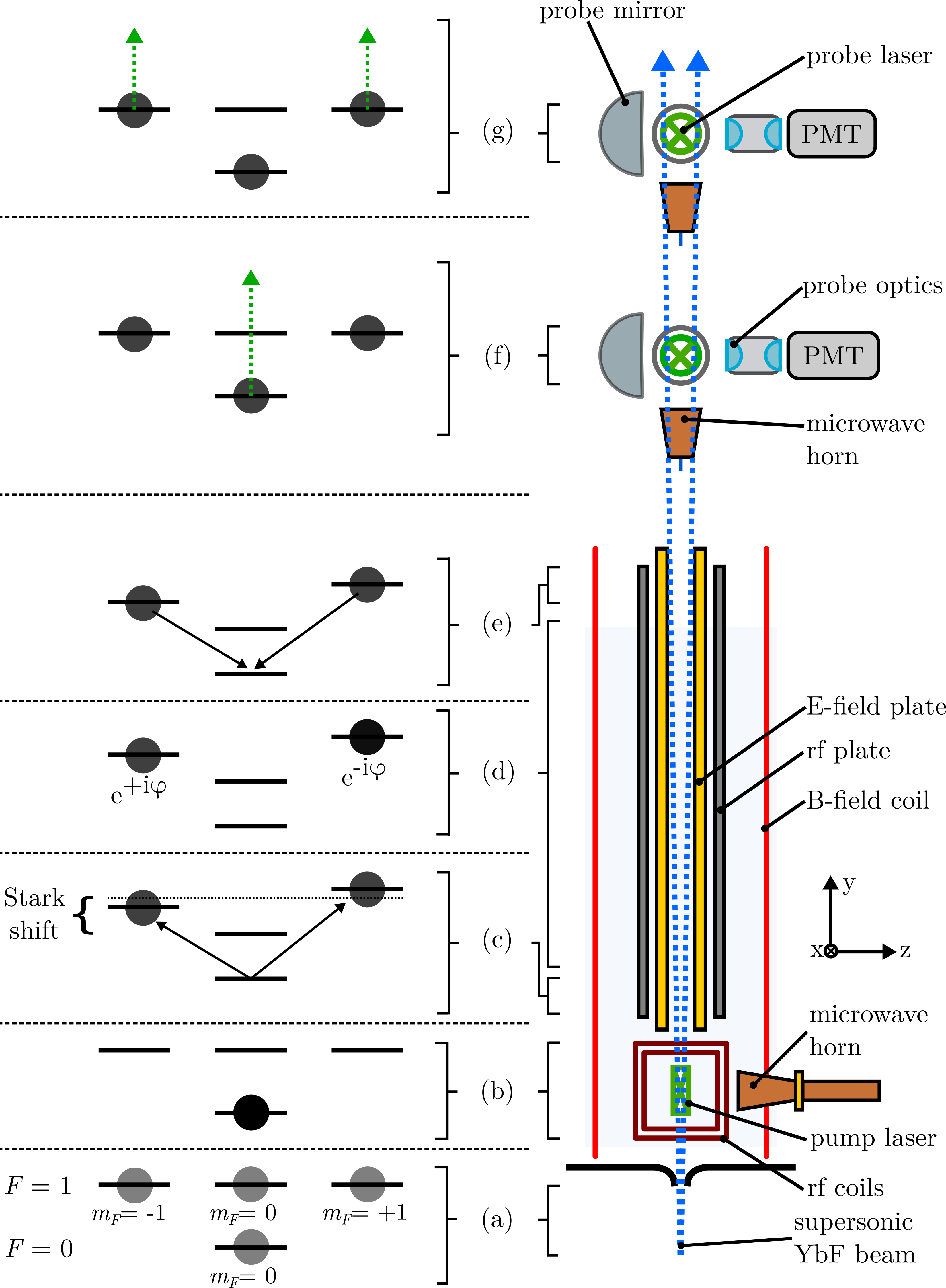}
    \caption{Overview of the experiment and interferometer states. (a) Pulsed beams of YbF molecules are produced with equal populations in the four $N=0$ sublevels. (b) Population is optically pumped using microwave, rf and optical fields into $F=0$. (c) An equal superposition of $m_F = +1$ and $m_F = -1$ is created by an rf pulse. (d) The two populated levels accumulate a relative phase due to interaction with $\bm{E}$ and $\bm{B}$ fields. (e) Population in the original superposition is projected back into $F=0$ by an rf pulse. (f) Population in $F=0$ is detected. (g) Population in $F=1$ is detected.}
    \label{fig:experimentOverview}
\end{figure}

An overview of the experiment is shown in Fig.~\ref{fig:experimentOverview}. A pulsed supersonic beam of YbF molecules, which has a forward velocity of \SI{600}{m/s} and a temperature of \SI{2}{K}, is produced by ablating a Yb target with a pulsed Nd:YAG laser into a supersonically expanding gas jet of Ar and  SF$_6$~\cite{Tarbutt2002}. The number of molecules participating in the experiment is increased by transferring population from the $N= 0,1,2$ states into the $(N,F) = (0,0)$ state by the new optical pumping scheme, which will be described in detail in Section 3. The molecules enter a region of uniform electric and magnetic fields, $\bm{E} = E\bm{\hat{z}}$ and $\bm{B} = B\bm{\hat{z}}$. In the electric field, the $m_F = 0$ sublevel of $F=1$ is shifted downwards relative to the $m_F = \pm 1$ states by the Stark interaction. The magnetic field breaks the degeneracy of the $m_F = \pm 1$ states via the linear Zeeman interaction, shifting the energies of the states by $g \mu_B m_F B $, where $g \simeq 1$. A non-zero eEDM, $d_e$, will further shift the energies of the states by $- d_e m_F E_\text{eff}$, where $E_\text{eff} = \eta E_\text{eff,max}$ and for YbF, $E_\text{eff,max} \approx \SI{-26}{GV \per cm}$~\cite{Kozlov1997}. The polarisation factor $\eta$ depends on the strength of the electric field applied. In our experiment, with $E = \SI{10}{kV \per cm}$, $\eta = 0.558$, so $E_\text{eff} \approx \SI{-14.5}{GV \per cm}$. A $\pi$-pulse of rf magnetic field, polarised along the $x$-axis, then transfers the population into an equal superposition of $m_F = +1$ and $m_F=-1$ states in $F=1$. The molecules evolve in the $\bm{E}$ and $\bm{B}$ fields for a time $\tau = \SI{800}{\mu s}$, and another rf $\pi$-pulse is applied which projects any population remaining in the original superposition back into the $F=0$ state. When the rf pulses are perfect $\pi$-pulses and exactly on resonance, the probability of finding a molecule in the $F=0$ or $F=1$ state is $p_0 = \cos^2{(\phi_B + \phi_E)}$ or $p_1 = \sin^2{(\phi_B + \phi_E)}$, where the phases $\phi_B = g\mu_B B \tau / \hbar$ and $\phi_E = - d_e E_\text{eff} \tau / \hbar$ are due to the Zeeman and eEDM interactions. The populations in $F=0$ and $F=1$ states are then sequentially measured by laser-induced fluorescence detection in two separate detectors labelled A and B. The detection scheme will be discussed in greater detail in Section~\ref{sec:stateDetection}.

We call the number of photons counted in each detector $s_\text{A}$ and $s_\text{B}$, and combine them to form a quantity called the asymmetry, defined as
\begin{equation}
\label{eq:asymmetryDefn}
    \mathcal{A} = \frac{s_\text{A}-s_\text{B}}{s_\text{A}+s_\text{B}}.
\end{equation}
This quantity is immune to shot-to-shot fluctuations in the number of molecules from the source, $N_\text{mol}$. In the ideal detection case, we have
\begin{eqnarray}
s_\text{A} &= N_\text{mol}\epsilon p_0 = N_\text{mol}\epsilon\cos^2\phi, \nonumber \\ s_\text{B} &= N_\text{mol}\epsilon p_1 = N_\text{mol}\epsilon\sin^2\phi,
\end{eqnarray}
where $\phi = \phi_B + \phi_E$ and $\epsilon$ is the number of photons detected per molecule, which we define to be the detection efficiency. The asymmetry is $\mathcal{A} = \cos 2 \phi$ in this case. We define the contrast, $\mathcal{C}$, as the amplitude of the $\cos{2 \phi}$ term in the asymmetry,
\begin{equation}
\label{eq:contrast}
    \mathcal{A} = \mathcal{C} \cos 2\phi.
\end{equation}
In the perfect experiment, we have $\mathcal{C} = 1$. We discuss imperfections which reduce $\mathcal{C}$ in Section~\ref{sec:contrast}.

The experiment measures $\phi_E$ by reversing the direction of the external electric field $E$ and measuring the resulting change in $\mathcal{A}$, from which we can extract $d_e$. The uncertainty of the measurement is therefore given by
\begin{eqnarray}
\label{eq:edmUncertainty}
    \sigma_{d_e} 
    &= \frac{\hbar}{E_\text{eff} \tau} \sigma_\phi \nonumber \\
    &= \frac{\hbar}{E_\text{eff} \tau} \left| \frac{\partial \phi}{\partial \mathcal{A}} \right| \sigma_\mathcal{A} \nonumber \\
    &= \frac{\hbar}{E_\text{eff} \tau} \frac{1}{2 \mathcal{C} | \sin 2 \phi |} \sigma_\mathcal{A}.
\end{eqnarray}

To minimise the uncertainty, we use the magnetic field $\bm{B}$ to set $\phi_B = \pi / 4$. At this phase, provided that $\epsilon \ll 1$, the uncertainty in the asymmetry can be shown to be limited by Poissonian statistics of photon-counting, $\sigma_\mathcal{A} = 1/\sqrt{s_\text{T}}$, where the total count $s_\text{T} = s_\text{A} + s_\text{B}$. We will elaborate on this in Section~\ref{sec:stateDetection}, as optical-cycling detection can lead to noise in excess of this simple shot-noise limit~\cite{Lasner2018}. In our experiment, we estimate the detection efficiency to be $\epsilon \approx 0.06$, so we can use the shot noise of the detected photons as that of the experiment. The shot-noise limited sensitivity of the experiment can then be written as
\begin{equation}
\label{eq:edmUncertaintyPiBy4}
    \xi_{d_e} = \frac{\hbar}{2 E_\text{eff} \tau \mathcal{C} \sqrt{s_\text{T}}}.
\end{equation}
This shows that to maximise the sensitivity of the experiment, we should maximise the coherence time, contrast and total photon count. The rest of the paper will focus on new techniques we have implemented to measure both $s_\text{A}$ and $s_\text{B}$, and to increase the count rate and contrast.

\section{State preparation}
\label{sec:statePrep}

\begin{figure}[tb]
\centering
\includegraphics{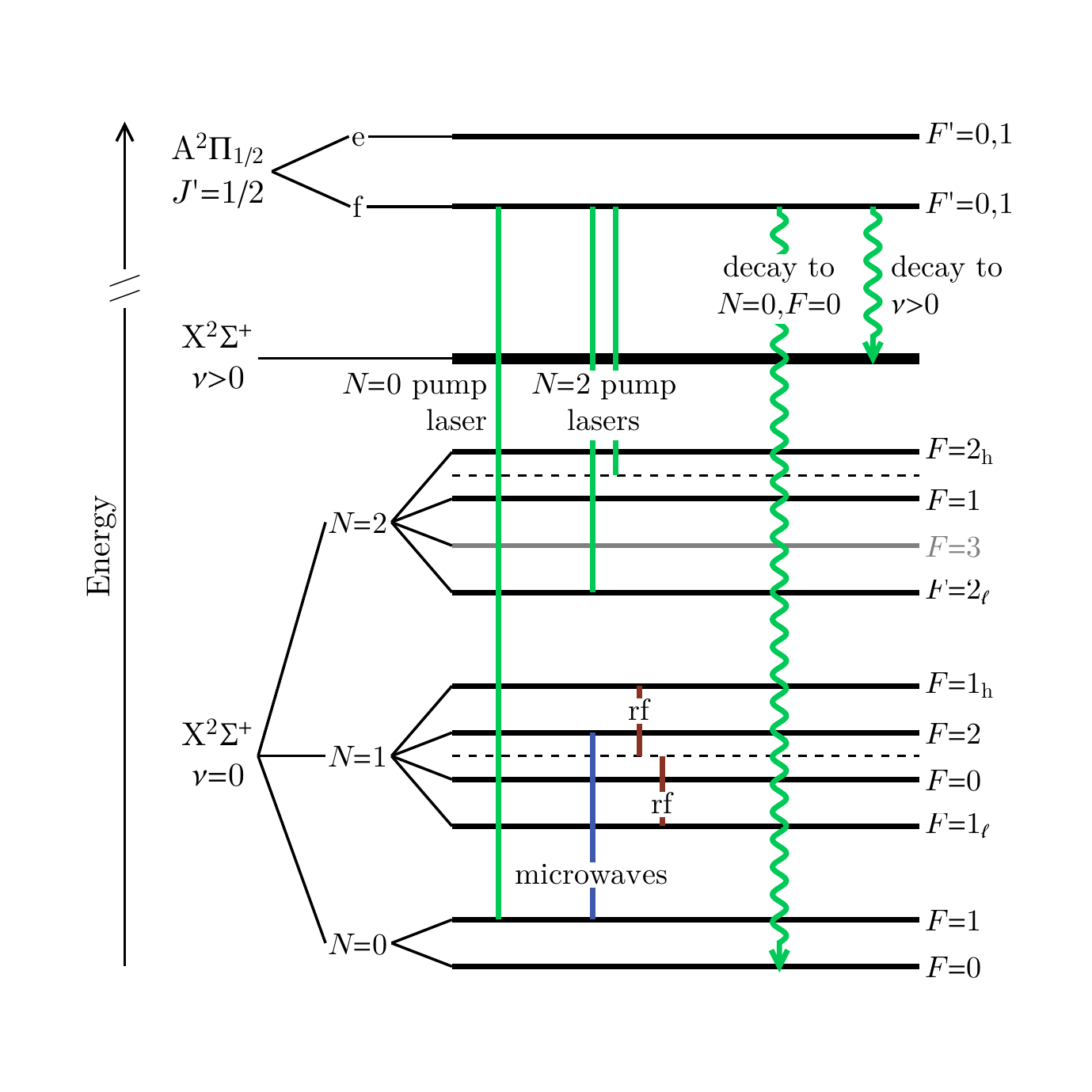}
\caption{Relevant transitions for state preparation. Population in $(N, F) = (0,1)$, $(2,1)$, $(2,2_\ell)$ and $(2,2_\text{h})$ are optically pumped into $(0,0)$ via the odd-parity $J' = 1/2$ state. Population in $N=1$ is added by coupling all its hyperfine states with $(0,1)$ using a combination of one microwave field and two rf fields.}
\label{fig:pumpingleveldiagram}
\end{figure}

Figure~\ref{fig:pumpingleveldiagram} shows the relevant transitions in YbF for the state preparation scheme. We want to increase the number of molecules in the $\text{X}^2\Sigma^+ (\nu = 0, N = 0, F=0)$ ground state, which is the starting point for the interferometer. Doing so increases $N_\text{mol}$ and therefore the total signal $s_\text{T}$.

We optically pump population from the even-parity states $(N,F) = (0,1)$, ($2,1$), ($2, 2_\ell$) and ($2, 2_\text{h}$) to $(0, 0)$ via the odd-parity $\text{A}^2\Pi_{1/2} (\nu'=0, J'=1/2, f)$ state. The ($2, 2_\text{h}$) and ($2,1$) states are closely spaced in energy, and are addressed by a single laser frequency. Due to parity and angular momentum selection rules, the only loss channel is spontaneous decay into higher vibrational states, $\nu > 0$, which has a branching ratio\footnote{The Franck-Condon factor for the $\text{A}(\nu = 0) \rightarrow \text{X}(\nu = 0)$ transition was measured to be 0.928~\cite{Zhuang2011}.} of 0.072. Population in ($2,3$) does not participate in the optical pumping scheme because the state has too much angular momentum. The $N=1$ states have odd parity, and therefore cannot be optically pumped into the even-parity state ($0, 0$). Instead, we couple all the hyperfine states in $N=1$ to the ($0, 1$) state using one microwave field and two rf fields, as shown in Fig.~\ref{fig:pumpingleveldiagram}, so that the population in $N=1$ can then be optically pumped into ($0,0$) as well.

\begin{figure}[t]
\centering
\includegraphics{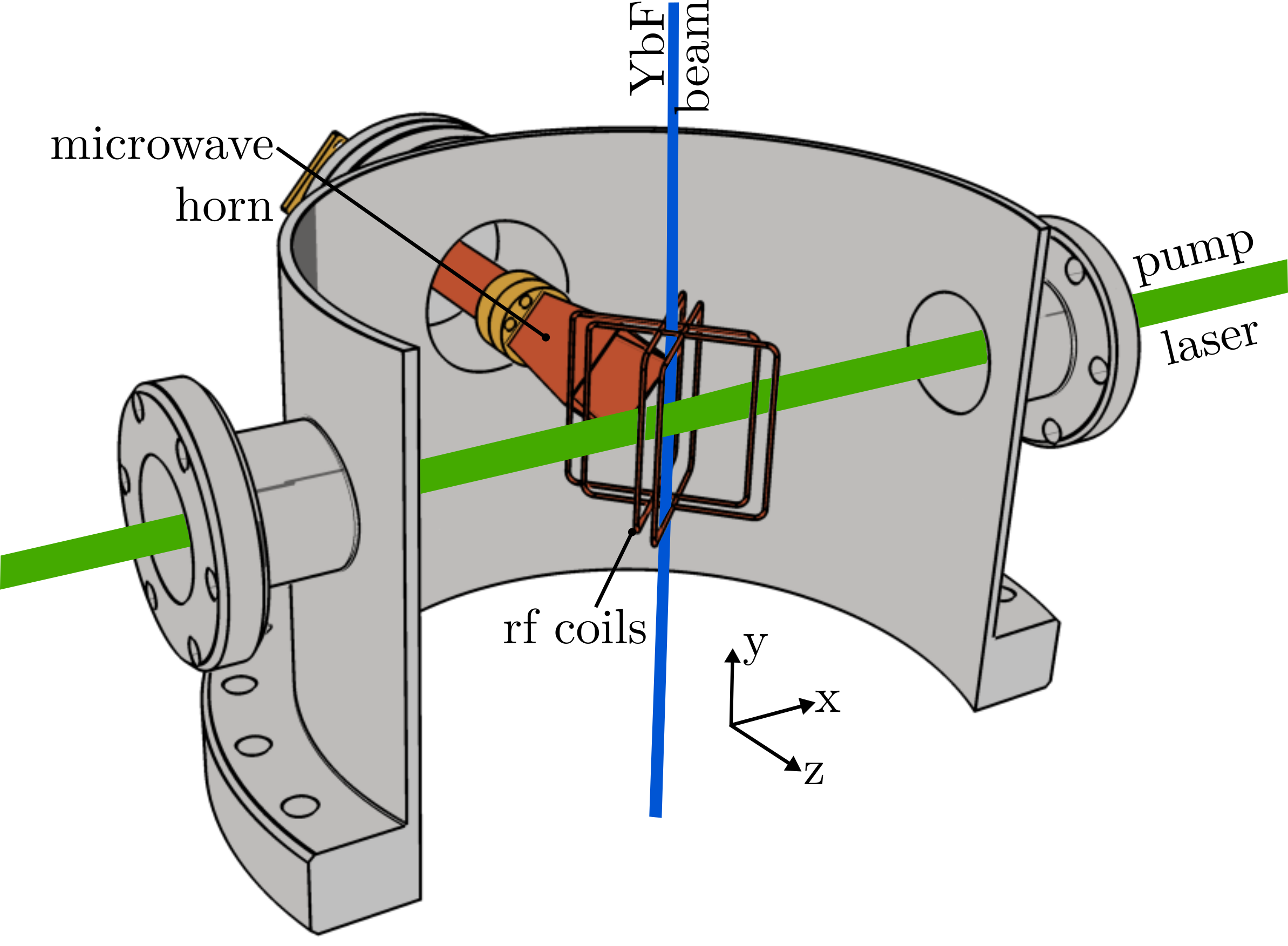}
\caption{The state preparation region. The radiation needed for the optical pumping scheme is provided by two counter-propagating laser beams, the microwave horn, and the rf coils. The rf coils are driven by a drive coil that is not shown. There is also a clean-up beam that co-propagates with one of the pump laser beams.}
\label{fig:pumpdrawing}
\end{figure}

Figure~\ref{fig:pumpdrawing} shows the state preparation region. The molecular beam travels vertically along the $y$-axis. The pump lasers are a set of counter-propagating elliptical beams which travel along the $x$-axis. These beams have $1/\mathrm{e}^2$ diameters of $23.5 \times \SI{2.2}{mm}$ along the $y$ and $z$ axes, providing an interaction time of $\SI{40}{\mu s}$ with the molecules. The peak laser intensities are approximately $\SI{120}{mW \per cm^2}$, $\SI{340}{mW \per cm^2}$ and $\SI{140}{mW \per cm^2}$ for the frequency components addressing the $(N,F)=(0,1)$ state, the ($2,2_\text{h}$) and ($2,1$) states, and the ($2,2_\ell$) state, respectively. There is also a clean-up beam, which co-propagates with one of the pump lasers in order to remove background signal in the detectors, as explained in Section~\ref{sec:contrast}.

A microwave horn delivers a maximum of \SI{19}{dBm} of power at the frequency resonant with the $(0,1) \rightarrow (1,2)$ transition. A set of resonant rf coils surround the molecular beam path, with resonances tuned to \SI{30.9}{MHz} and \SI{161.2}{MHz}. The two rf fields are first generated by voltage-controlled oscillators, combined and amplified, then sent into the machine to a drive coil which couples the rf to the resonant coils. The excited resonant coils then generate rf radiation which drive M1 transitions between the hyperfine levels in $N=1$. Both the microwave and rf powers are optimised empirically to maximise the population in ($0,0$).

\begin{figure}[tb]
\centering
\includegraphics{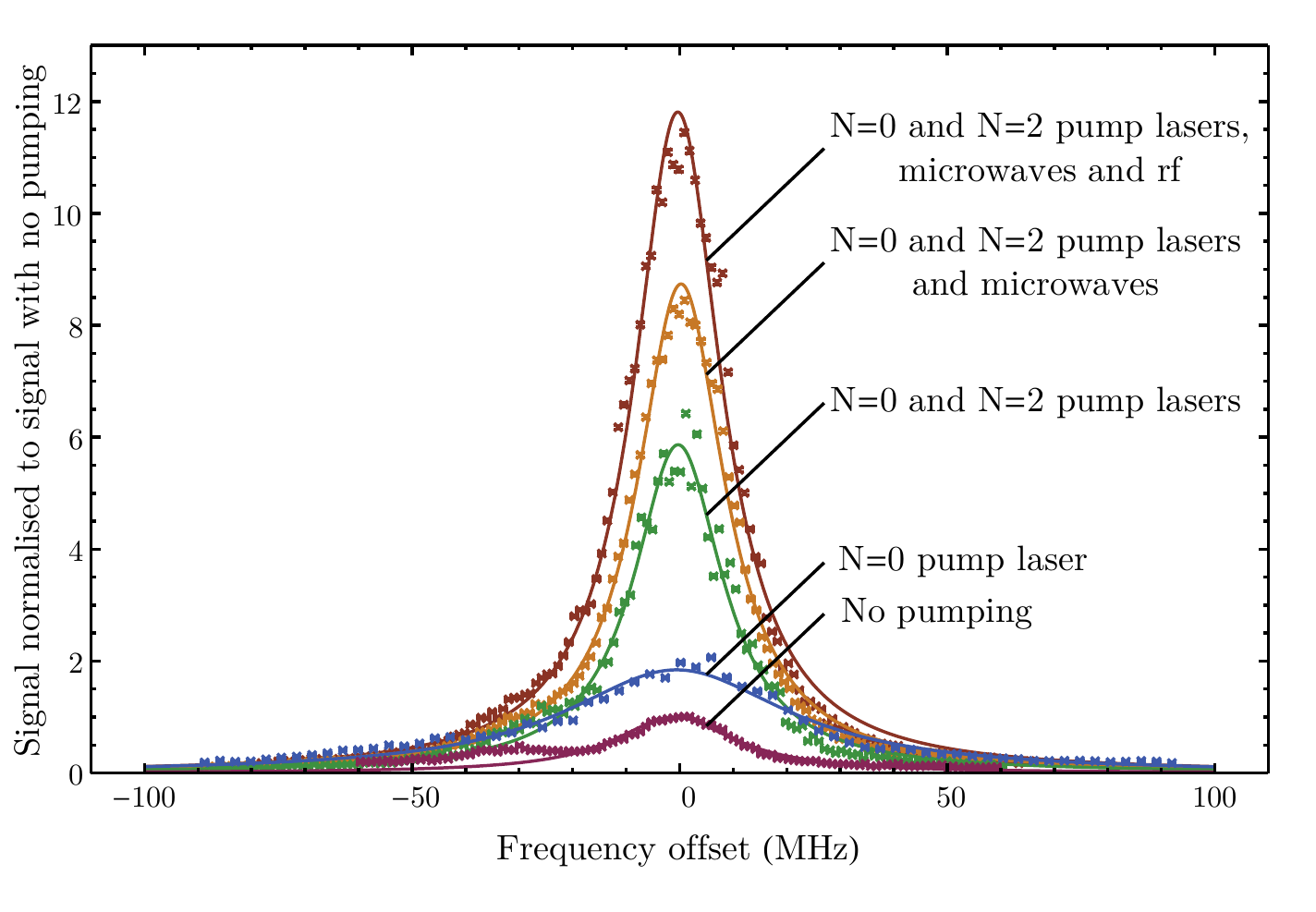}
\caption{Increase in molecule population in the $(N,F)=(0,0)$ state due to optical, microwave and rf fields. The signal is normalised to the $F=0$ signal with no optical pumping.}
\label{fig:improvementsfrompumping}
\end{figure}

Figure~\ref{fig:improvementsfrompumping} shows the results of the state preparation scheme. Using one of the two detectors, we measure the population in $(N,F)=(0,0)$ by scanning the frequency of a weak probe laser around the $F=0$ component of the Q(0) transition, and detecting the laser-induced fluorescence. When we add only the $N=0$ pump laser, which was the method used in the previous eEDM measurement \cite{Hudson2011,Kara2012}, the signal increases by a factor $1.8 \pm 0.2$ over the signal from the thermal population in $F=0$. The origin of the error bar is shot-to-shot fluctuations in the number of molecules. Numerical simulations using OBEs~\cite{Devlin2015} predict a factor of 1.9, in good agreement with the measurement. When we add the $N=2$ pump laser, the enhancement factor grows to $5.9 \pm 0.6$. Adding the microwave field increases this to $8.7 \pm 0.8$, and adding the rf fields increases this further to $11.8 \pm 1.2$. This last factor is again in agreement with OBE simulations which predict a signal increase of 10.5.

\section{State detection}
\label{sec:stateDetection}

A simple way to measure the population in one of the two hyperfine states is to excite molecules in that state and detect the resulting laser-induced fluorescence using the Q(0) transition (see Fig.~\ref{fig:energyLevelDiagram}). This method, which we call Q(0) detection, is the one used in the previous YbF eEDM measurement. It has two disadvantages. First, there is no closed transition from $N=0$, so each molecule scatters just a small number of photons before it is optically pumped to a different state. From OBE simulations, we calculate that an $(N,F) = (0,0)$ molecule scatters only 1.2 photons on average. Second, the population in one of the two hyperfine sublevels is partially pumped into the other, so the measurement of one population changes the other.

\begin{figure}[tb]
\centering
    \includegraphics{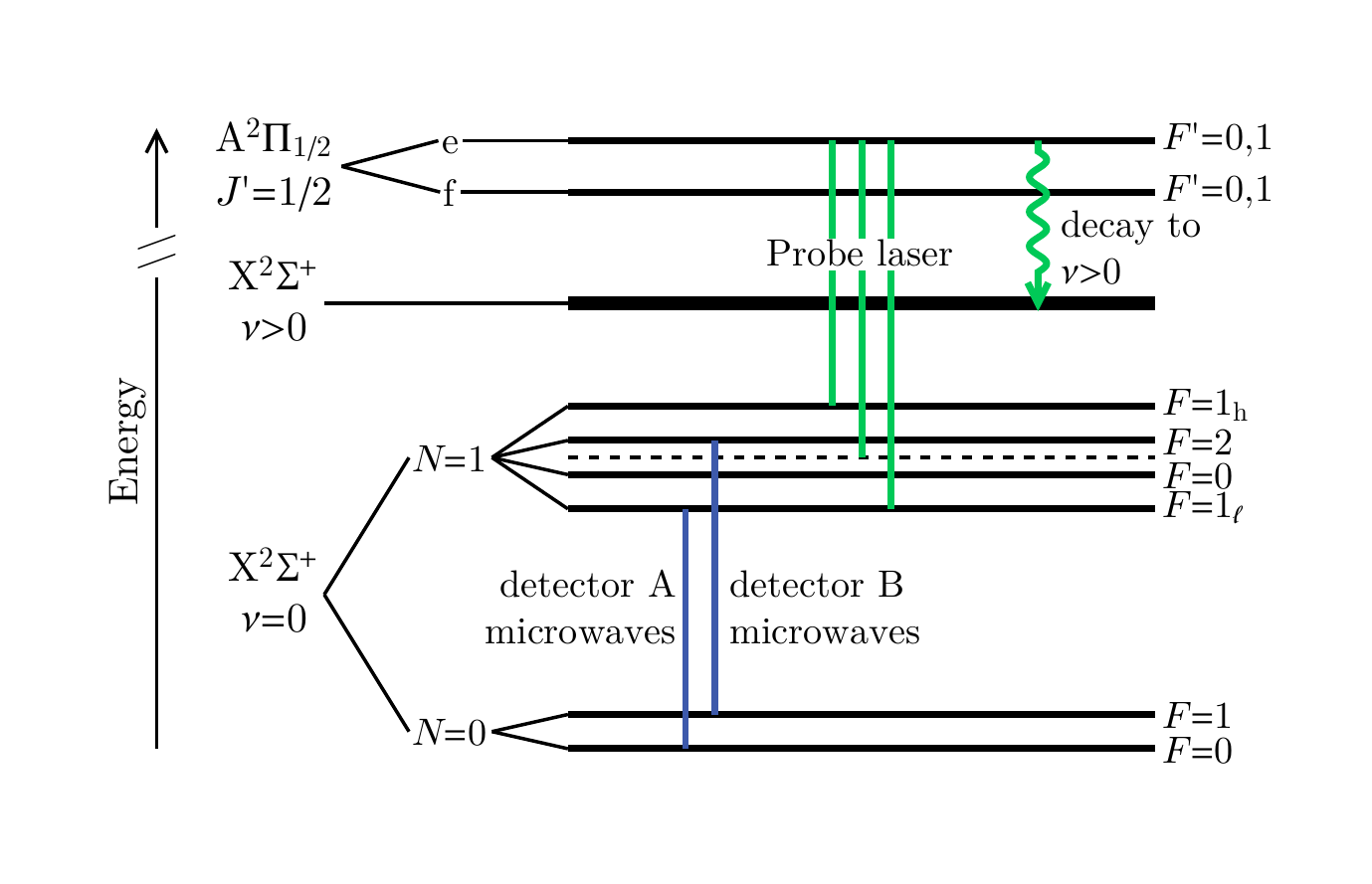}
\caption{Level diagram showing the relevant transitions for the state detection scheme.}
\label{fig:probeleveldiagram}
\end{figure}

Here we implement a better detection method using the P(1) transition which scatters many more photons and also allows us to measure $s_\text{A}$ and $s_\text{B}$ independently. Our P(1) detection scheme is shown in detail in Fig.~\ref{fig:probeleveldiagram}. We detect the molecules in $F=0$ and $F=1$ sequentially, in detectors A and B. In detector A, a resonant microwave field couples the states $(N, F) = (0, 0)$ and $(1, 1_\ell)$. A probe laser together with sidebands, tuned to the the P(1) transition, is used to detect the molecules. In detector B, the same is carried out except with resonant microwaves coupling the states $(0, 1)$ and $(1, 2)$ instead, which detects population in $F=1$. The P(1) transition is rotationally closed: molecules can only either decay back to the $N=1$ states in the $\nu = 0$ manifold, or to higher vibrational states. The branching ratio for the decay to $\nu > 0$ states is $\Gamma_{\nu > 0} = 0.072$, so each molecule scatters, on average, $1/(1-\Gamma_{\nu > 0}) = 13.8$ photons before it becomes dark to the probe laser.

\begin{figure}[tbp]
\centering
\includegraphics{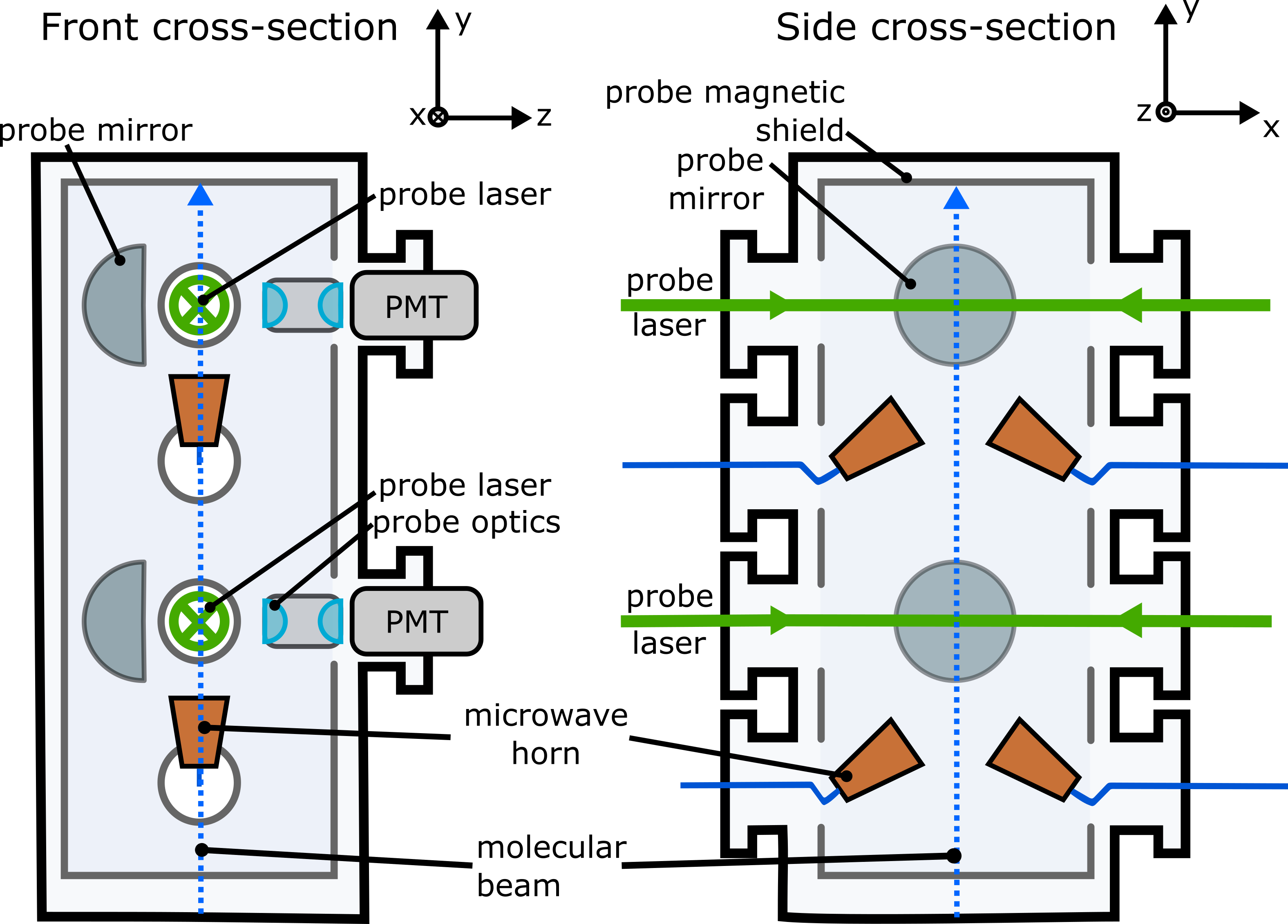}
\caption{The detection chamber with two detection regions.}
\label{fig:probedrawing}
\end{figure}

The two detectors are illustrated in Fig.~\ref{fig:probedrawing}. Each includes two microwave horns, although only one is used when detecting the molecules. The other horn is included so that the standing wave pattern of the microwave field can be changed, allowing us to test for systematic effects in the experiment. In each detector, we can apply either microwave frequency, so the order of detecting the $F=0$ and $F=1$ populations can be reversed to check for systematic effects. An optical fibre delivers light of all three probe frequencies to the upper detector and this same light is routed to pass through the lower detector to ensure some common-mode rejection of fluctuations in the light. A second fibre delivers another probe beam to the lower detector, which counter propagates along the same path. Each beam has a circular Gaussian profile and a $1/\text{e}^2$-radius of \SI{6}{mm}. The combined peak intensities of the two beams are \SI{80}{mW \per cm^2}, \SI{160}{mW \per cm^2} and \SI{80}{mW \per cm^2} for the three different frequencies which address the $F=1_\ell$, $F=0/F=2$, and $F=1_\text{h}$ hyperfine levels respectively, as shown in Fig.~\ref{fig:probeleveldiagram}. The middle frequency addresses two hyperfine levels as they are very closely spaced in energy (see Fig.~\ref{fig:energyLevelDiagram}).

The resulting laser-induced fluorescence from the molecules is collected by a large spherical mirror and imaged onto a PMT by a pair of aspheric lenses. With such a high laser power in the detection region, it is imperative that we minimise the background signal due to laser scatter. Although this background can be measured\footnote{This is done by measuring the PMT signal when the molecules are not in the detection region.} and subtracted away, there is still the increase in uncertainty due to the noise in the laser scatter. We minimise this background by blackening the interior surfaces of the chamber\footnote{Some surfaces were blackened using soot from an acetylene flame, and others were blackened using a black paint from Alion Science (MH2200).}, using optical baffles, extending the input and output arms of the laser ports, and angling the port windows. With these measures, the scatter comes down to about 50 (100) photons in 1 $\mu s$ in the lower (upper) detection region, equivalent to 5\% (10\%) of the molecular signal averaged over the time window used in the data analysis. The increase in uncertainty in the measurement due to noise in the laser scatter is then negligible.

\begin{figure}[tb]
    \centering
    \includegraphics{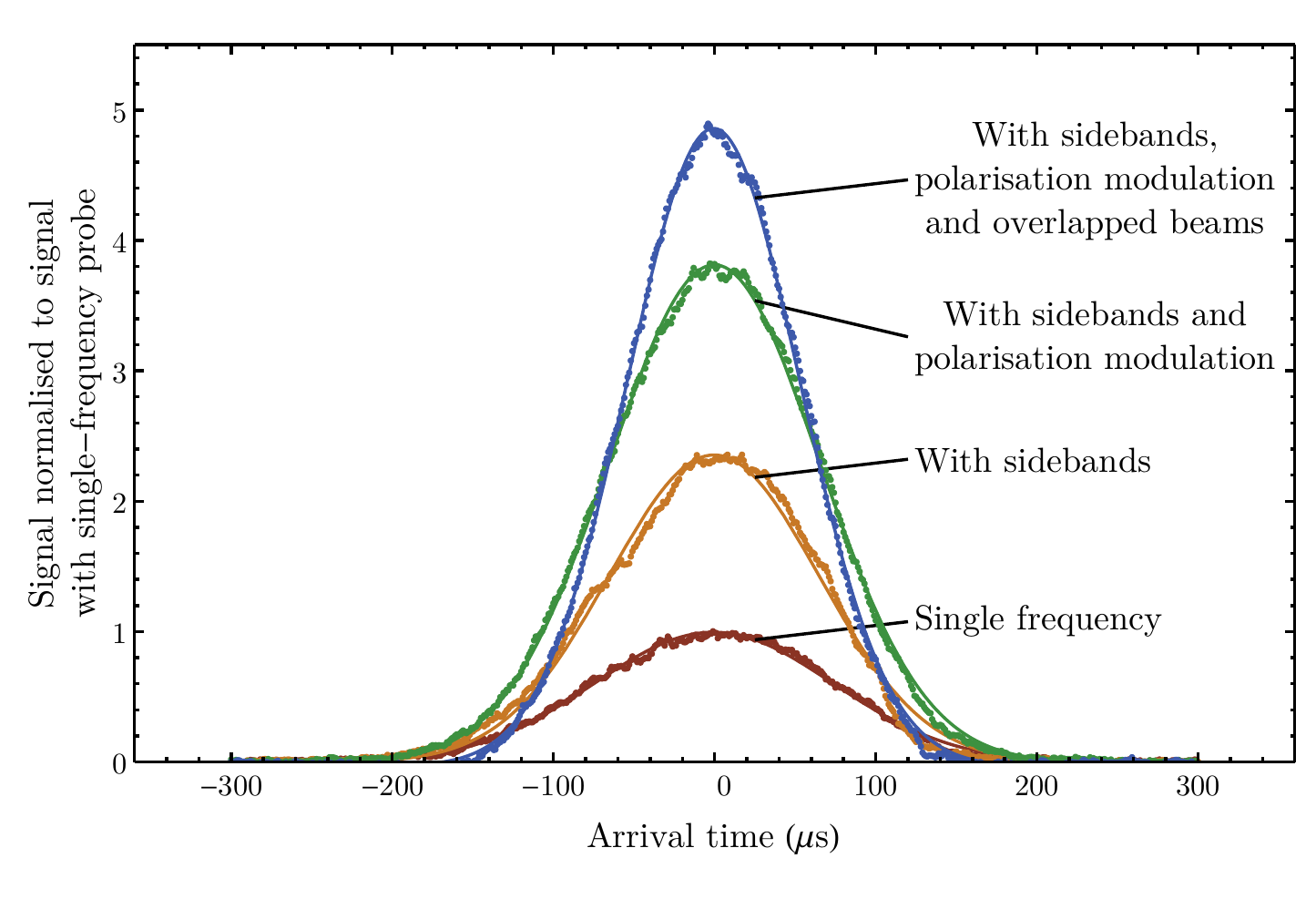}
    \caption{Increase in the detected signal using P(1) detection due to methods that improve the optical cycling of the molecules. The signals are normalised to the peak signal when only a single frequency is used in the probe laser and the backgrounds are subtracted.}
    \label{fig:detectionEffImprovements}
\end{figure}

To ensure that the optical cycling due to the rotationally-closed nature of the P(1) transition is maximised, we use three probe laser frequencies (see Fig.~\ref{fig:probeleveldiagram}) so that we excite molecules from all of the $N=1$ hyperfine levels. Furthermore, some of the P(1) transitions have $F \geq F'$, where $F, F'$ are the total angular momenta of the hyperfine states in the ground and excited electronic states. It is well known that if the driving laser has a static polarisation, such a configuration has dark states, which will significantly reduce the number of photons scattered per molecule~\cite{Berkeland2002}. We destabilise these dark states by modulating the polarisation of the probe beams between $y$ and $z$ (with circular in between) using an electro-optic modulator (EOM). We use a rate of \SI{0.9}{MHz} because our OBE simulations give a very broad optimum centred on this frequency. We also counter-propagate the probe beams in order to create polarisation gradients in the $x$-direction. The molecules then experience a further varying laser polarisation as they fly through the detection region since they have non-zero transverse velocity. 

Figure~\ref{fig:detectionEffImprovements} shows the improvement in the signal from one of the detectors when we use the methods described above to increase the optical cycling of the molecules. In all the cases, the total laser power is \SI{100}{mW}. For the counter-propagating case, we have used \SI{50}{mW} per beam. The results are summarised in Table~\ref{tab:signalIncreaseComparison}. When sidebands are added to the probe laser, the signal detected increases by a factor of $2.3 \pm 0.3$, in agreement with our simulations. When we also modulate the polarisation of the probe lasers, the signal detected is a factor of $3.8 \pm 0.4$ higher than the single-frequency case. This is lower than the OBE simulation prediction of 5.2, suggesting that the polarisation of the probe light is not being fully modulated after passing through the optical fibres. If we counter-propagate (keeping the total power constant), the increase in signal is $4.9 \pm 0.7$ times, which agrees with the predicted increase when the dark states are fully destabilised. The OBE simulations additionally show that even with full polarisation modulation, we have yet to reach the maximum possible average of 13.8 photons per molecule. This is due to insufficient interaction time of the molecules with the probe lasers in the detection region. A larger probe beam diameter could have been used, but it would have resulted in an unacceptably large amount of laser scatter in our detection chamber, and also made it more challenging to image all the signal photons onto the PMTs.

\Table{\label{tab:signalIncreaseComparison} Comparison of signal increase due to detection techniques with that predicted by OBE simulations.}
\br

         & & & Average number of \\
         & Signal & Signal & photons scattered \\
         & increase & increase & per molecule \\
         Detection method & (experiment) & (simulation) & (simulation) \\
        \mr
        Single-frequency probe laser & 1 & 1 & 2.3 \\
        Probe laser with sidebands & $2.3 \pm 0.3$ & 2.0 & 4.6 \\
        Sidebands and & \multirow{2}{*}{$3.8 \pm 0.4$} & \multirow{2}{*}{5.2} & \multirow{2}{*}{11.9} \\
        polarisation modulation & & & \\
        Sidebands, polarisation& \multirow{3}{*}{$4.9 \pm 0.7$} & \multirow{3}{*}{5.2} & \multirow{3}{*}{11.9} \\
        modulation and & & & \\
        counter-propagating beams & & & \\
\br
\endTable

Lasner and DeMille have carefully considered the signal-to-noise ratio obtained in an experiment such as ours, where each molecule scatters many photons before being pumped to a dark state, and each photon has a certain probability of being detected~\cite{Lasner2018}. They find that the expression for the shot noise has to be corrected because the probability distribution of the detected photons per molecule is broader than that of a Poissonian distribution. In our case, this correction is small because the number of photons detected per molecule is much smaller than 1. If the detection efficiency can be improved in the future, it will be important to take proper account of the excess noise factor discussed in Ref.~\cite{Lasner2018}.

To conclude this section, the average of 11.9 photons scattered per molecule using this P(1) detection is an order of magnitude higher than the 1.2 photons scattered per molecule using the previous Q(0) detection method. Moreover, P(1) detection allows us to detect both $F=0$ and $F=1$ molecules independently, which doubles the number of useful molecules in the experiment compared to Q(0) detection, further improving our sensitivity. The next section discusses imperfections in our P(1) detection scheme and describes methods to overcome them.

\section{Improving the contrast}
\label{sec:contrast}

We have shown in Section~\ref{sec:method} that in the ideal case, the photon counts in detectors A and B are given by $s_\text{A} = N_\text{mol} \epsilon p_0 = N_\text{mol} \epsilon \cos^2\phi$ and $s_\text{B} = N_\text{mol} \epsilon p_1 = N_\text{mol} \epsilon \sin^2\phi$. If there are imperfections in the two rf $\pi$-pulses, due to imperfect pulse area or frequency, then the probability of detecting a molecule in $F=0$ and $F=1$ after the second pulse has the form~\cite{Kara2012}
\begin{eqnarray}
\label{eq:moleculepopulations}
p_0 =&\ a_I \cos^2{\phi} + a_R\cos{\phi} + a_{C,0}, \nonumber \\
p_1 =&\ a_I \sin^2{\phi} - a_R\cos{\phi} + a_{C,1},
\end{eqnarray}
where $a_I$ is the amplitude of the main interference signal we want to measure, $a_R$ is the amplitude of the Ramsey fringe that arises from the residual coherence between $F=0$ and $F=1$, and $a_{C,i}$ is a constant term that is different for the two states. The amplitudes are constrained such that $p_0 \geq 0$, $p_1 \geq 0$, and $p_0 + p_1 = 1$.

Now let us consider imperfections in the detection scheme and how they affect the signals. Let the lower detector be A, i.e. the one that detects population in $F=0$. First, the optical cycling process does not completely deplete the population in $N=1$, and so we write the fraction of population left in $N=1$ after detector A as $f_\text{left}$. Next, we include the observation that there is some off-resonant driving of the $F=1$ state when driving the $F=0$ microwave transition, and vice-versa, and denote the fraction of population that is off-resonantly driven as $f_\text{or}$. We account for the difference in detection efficiencies of the two regions by a fraction $f_\epsilon$, so that the detection efficiencies are $(1\pm f_\epsilon) \epsilon$ in the two regions. In addition to the background due to laser scatter from the apparatus, which we have discussed earlier, we have another constant background term due to scattering from molecules in other states, which do not participate in the experiment. We represent the latter as a fraction of the total number of molecules, $f_\text{bg,A/B}$. The measured photon counts can then be written as
\begin{eqnarray}
\label{eq:detectedSignalsImperfect}
s_\text{A} =&\ N_\text{mol} \epsilon a_I (1+f_\epsilon) (\cos^2{\phi} + f_\text{R} \cos{\phi} + f_\text{or} \sin^2{\phi} + f_\text{bg,A}), \nonumber \\
s_\text{B} =&\ N_\text{mol} \epsilon a_I (1-f_\epsilon) ( (1-f_\text{or}) (\sin^2{\phi} - f_\text{R} \cos{\phi})+ f_\text{left} \cos^2{\phi} + f_\text{bg,B}),
\end{eqnarray}
where we have omitted terms with products of small fractions. Here $f_R = a_R/a_I$ is the Ramsey term amplitude as a fraction of the signal amplitude. The asymmetry, as defined in Eq.~\ref{eq:asymmetryDefn}, can now be written as
\begin{eqnarray}
\label{eq:asymmetryImperfect}
\mathcal{A} &= \frac{s_\text{A}-s_\text{B}}{s_\text{A}+s_\text{B}} \nonumber \\
&\approx \left( \frac{f_\epsilon}{2} + f_\text{or} - \frac{3f_\text{left}}{4} + f_\text{bg,A} - f_\text{bg,B} \right) 
+ 2 f_R \cos{\phi} \nonumber \\
&\ + \left( 1 - f_\text{or} - f_\text{left} - f_\text{bg,A} - f_\text{bg,B} \right) \cos{2\phi} 
- \frac{1}{4} \left( 2 f_\epsilon + f_\text{left} \right) \cos{4\phi} \nonumber \\
&\equiv \mathcal{A}_c + \mathcal{A}_R \cos{\phi} + \mathcal{C} \cos{2\phi} + \mathcal{A}_d \cos{4\phi},
\end{eqnarray}
where again we have neglected all terms proportional to a product of two or more small quantities. If the order of detecting the molecules is reversed, we obtain a very similar expression to Eq.~\ref{eq:asymmetryImperfect}.

In line with our earlier definition of the contrast in Eq.~\ref{eq:contrast}, we have set $\mathcal{C}$ to be the coefficient of the $\cos 2 \phi$ term in the expression for $\mathcal{A}$. In addition, there is a constant term $\mathcal{A}_c$, a Ramsey term with amplitude $\mathcal{A}_R$ and a frequency-doubled term with amplitude $\mathcal{A}_d$. The Ramsey component can be removed using the methods described in Ref.~\cite{Kara2012}, and since all the other imperfections are small, the $\cos{2\phi}$ term dominates. The sensitivity of the experiment depends on how well we can measure a change in $\mathcal{A}$ from a corresponding change in the phase $\phi$, which is given by
\begin{eqnarray}
    \label{eq:asymmetrySlope}
    \left| \frac{\partial \mathcal{A}}{\partial \phi} \right|_{\phi = \pi/4} 
    &= \left\vert \left( -2\mathcal{C}\sin 2\phi -4\mathcal{A}_d \sin 4\phi \right) \right|_{\phi = \pi/4} = 2\mathcal{C} \nonumber \\
    &= 2 \left( 1 - f_\text{or} - f_\text{left} - f_\text{bg,A} - f_\text{bg,B} \right),
\end{eqnarray}
where we have set $\mathcal{A}_R = 0$ and $\phi$ is set to $\pi / 4$ to maximise the slope, as in Section~\ref{sec:method}. The sensitivity of the experiment can therefore be maximised by minimising the imperfection terms. We now look at each of these terms.

We minimise the off-resonant driving from the microwaves by optimising the detection microwave powers to achieve the best contrast. The fraction of off-resonant population after optimisation is measured to be $f_\text{or} = 0.04 \pm 0.01$. To minimise the leftover population, we destabilise the dark states in the $N=1$ levels by modulating the probe laser polarisation and counter-propagating the beams, as discussed in Section~\ref{sec:stateDetection}. After applying these methods, we measure $f_\text{left} = 0.17 \pm 0.02$.

\begin{figure}[tb]
    \centering
    \includegraphics{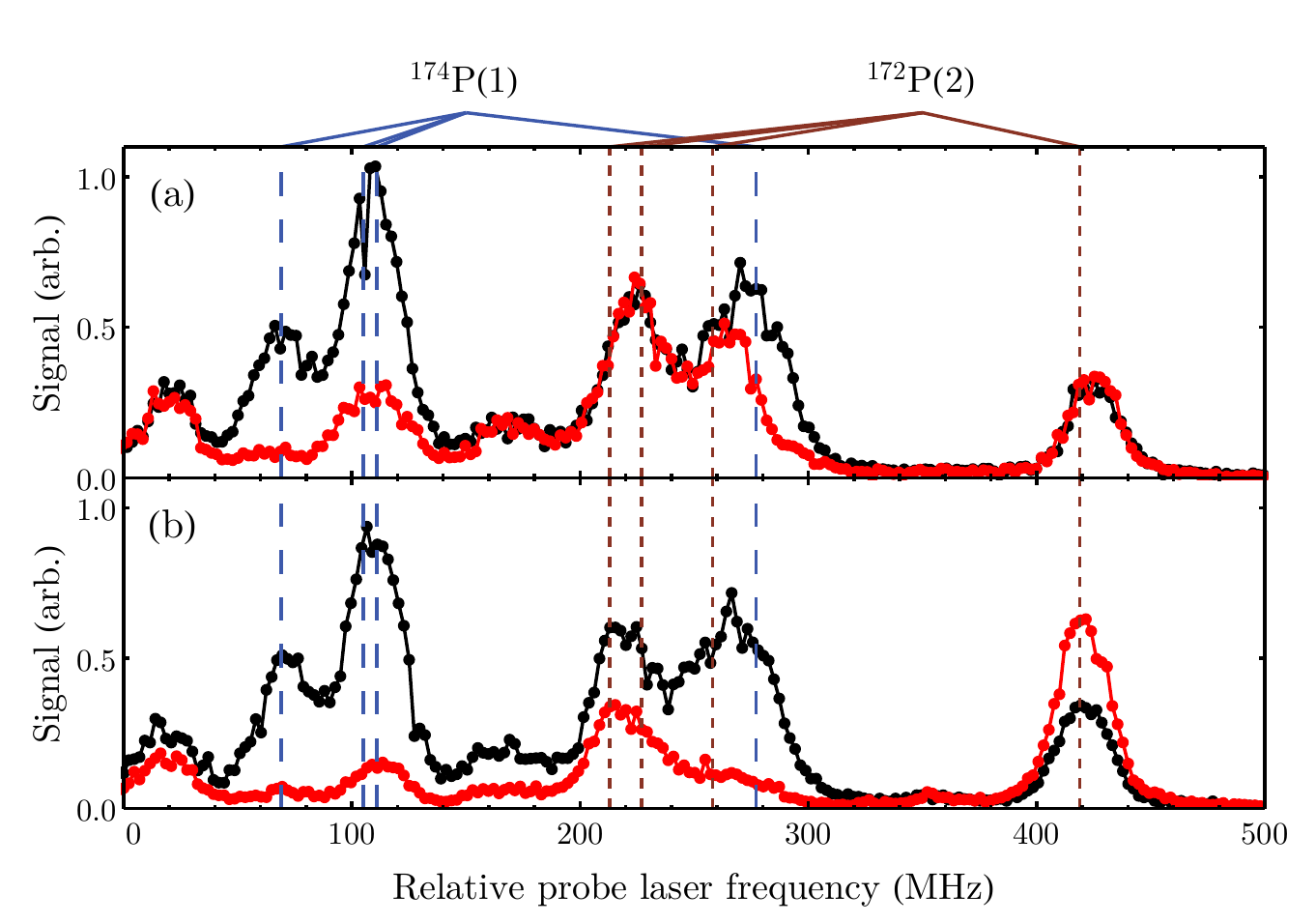}
    \caption{Spectral lines around the $^{174}\text{P}(1)$ transitions, obtained by using a weak probe laser ($\SI{400}{\mu W}$) in the detection region. The $^{174}\text{P}(1)$ and $^{172}\text{P}(2)$ transitions are indicated by the dashed lines. Black: no optical pumping was carried out. Red: (a) regular optical pumping is done, as in Section~\ref{sec:statePrep}; (b) The detection light ($^{174}\text{P}(1)$) was used to pump the molecules, in place of the regular optical pumping, in order to show the origin of the background signal in the detection region.}
    \label{fig:cleanUpSpectroscopy}
\end{figure}

The background scatter from the molecular beam mainly comes from two sources. The first source is molecular population in $N=1$ that remained after the optical pumping step. We show this in Fig.~\ref{fig:cleanUpSpectroscopy}(a), where we operated the usual optical pumping scheme, but scanned a weak probe laser in the detection region around the $^{174}\text{P}(1)$ transitions\footnote{Here, we use the superscript to denote the relevant isotope of Yb that the transition refers to.}. After optical pumping, there remains a significant amount of population in the $F=0/F=2$ hyperfine levels of $N=1$ in $^{174}$YbF. About 17\% of the population is left in $N=1$ after optical pumping.

\Table{\label{tab:contrast} Improvements in contrast. We show the contrast for the cases where either $F=0$ or $F=1$ population is detected first. Unless stated, the total laser power is \SI{100}{mW} (if using counter-propagating beams, this corresponds to \SI{50}{mW} per beam).}
\br

	\multirow{2}{*} & Contrast & Contrast \\
        & ($F=0$ first) & ($F=1$ first) \\
        \mr
        No improvements & 0.35 & 0.37 \\
        Clean-up beam (CU) & 0.44 & 0.45 \\
        Counter-propagating beams (CP) & 0.39 & 0.40 \\
        Polarisation modulation (PM) & 0.45 & 0.43 \\
        PM + CU & 0.56 & 0.53\\
        PM + CP + CU & 0.58 & 0.55\\
        PM + CP + CU at \SI{80}{mW} per beam & 0.62 & 0.61\\

\br
\endTable

The other isotopologues of YbF are the second source of background from molecular scattering. We show this in Fig.~\ref{fig:cleanUpSpectroscopy}(b). In the figure, we note that the $^{174}\text{P}(1)$ lines overlap with the $^{172}\text{P}(2)$ lines, so the latter can contribute to background scatter since our detection method requires high laser power, leading to very large power broadening. The natural abundances of $^{174}$YbF and $^{172}$YbF are 32\% and 22\% respectively, so this is a significant source of background. We demonstrate this by applying a pump laser tuned to the $^{174}\text{P}(1)$ transitions (the same laser frequencies as the probe lasers). The $N=1$ levels are depleted, though not completely, and there are other peaks in the vicinity that are also pumped out. These other peaks contribute to the background we observe in the detectors.

This background due to other molecules can be minimised by introducing some ``clean-up'' light into the optical pumping region. This clean-up light is obtained by picking off a small amount of probe laser light, broadening its frequency spectrum with an overdriven EOM such that its bandwidth is about \SI{200}{MHz}, and directing this light into the optical pumping region. Since the $^{172}\text{P}(2)$ transitions are not rotationally-closed, whereas the $^{174}\text{P}(1)$ transitions are, we remove the background without adversely affecting the efficiency of the regular optical pumping scheme. With the clean-up beam in place, we measure the contributions of the background signal (from both sources) to be $f_\text{bg,A} = 0.17 \pm 0.02$ and $f_\text{bg,B} = 0.06 \pm 0.01$.

\begin{figure}[bt]
\centering
\includegraphics{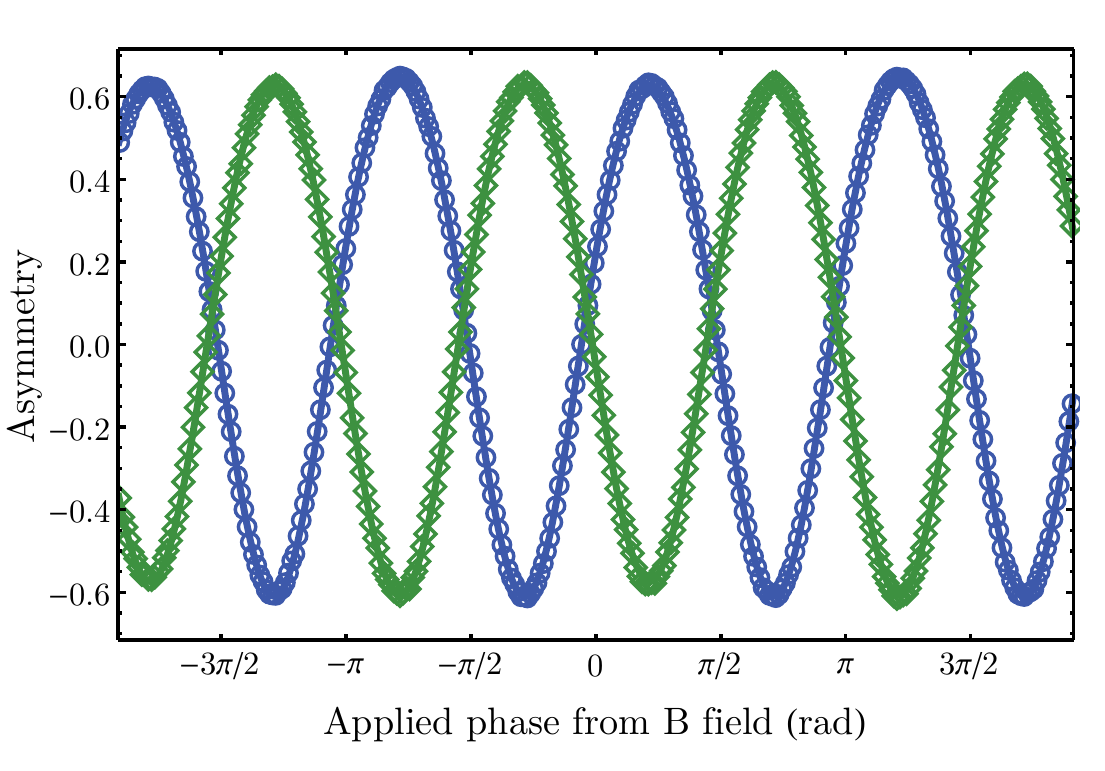}
\caption{Interference curves obtained by detecting population in $F=0$ first, then $F=1$ (blue open circles), and the other way round (green open diamonds). Solid lines are fits to Eq.~\ref{eq:asymmetryImperfect}.}
\label{fig:asymmetrycurve}
\end{figure}

Based on our measurements of the parameters appearing in Eq.~\ref{eq:asymmetrySlope}, we expect the contrast to be $| \mathcal{C} | = 0.56 \pm 0.03$. In Table~\ref{tab:contrast}, we show the improvement in contrast after applying the methods described in this section:
\begin{enumerate*}[label=(\roman*)]
    \item polarisation modulation of the probe lasers (PM),
    \item counter-propagating beams in the detector (CP) and
    \item clean-up beam in the optical pumping region (CU).
\end{enumerate*}
The contrast was measured by fitting interference curves, obtained by scanning the magnetic field, to Eq.~\ref{eq:asymmetryImperfect}. We have achieved a contrast of $|\mathcal{C}| = 0.62$ for the detector configuration where we measure $F=0$ population first, and a contrast of $|\mathcal{C}| = 0.61$ for the opposite configuration that measures $F=1$ population first, which is consistent with our estimation. Figure~\ref{fig:asymmetrycurve} shows the interference curves and fits with all the improvements and maximum laser power. We have added a phase offset $\phi_b$ to account for the ambient background magnetic field. The fit parameters are shown in Table~\ref{tab:fitParams}. The other terms in the asymmetry, $\mathcal{A}_c$, $\mathcal{A}_R$ and $\mathcal{A}_d$, are small compared to the contrast.

\Table{\label{tab:fitParams} Fit parameters for the data shown in Fig.~\ref{fig:asymmetrycurve}.}
\br

	Measuring & $\mathcal{C}$ & $\mathcal{A}_c$ & $\mathcal{A}_R$ & $\mathcal{A}_d$ & $\mathcal{\phi}_b$ (rad) \\
        \mr
        $F=0$ population first & 0.623 & 0.054 & -0.012 & -0.041 & -0.21~$\pi$ \\
        $F=1$ population first & -0.609 & 0.045 & 0.001 & -0.026 & 0.29~$\pi$ \\

\br
\endTable

\section{Sensitivity of the experiment}\label{sec:sensitivity}
We now investigate the statistical sensitivity for an eEDM measurement when we use the new techniques described above. We call a single pulse of the molecular beam a ``shot''. From each shot, we obtain two pulses of photon counts, $s_\text{A}(t)$ and $s_\text{B}(t)$, at the two detectors. The lower and upper detectors are at distances $L_1$ and $L_2$ from the source. The molecules reaching the upper detector at time $t$ arrive in the lower detector at the earlier time $t' = (L_1/L_2)t$, so we calculate the asymmetry as $\mathcal{A}(t) = (s_\text{A}(t') - s_\text{B}(t))/(s_\text{A}(t') + s_\text{B}(t))$. The function $\mathcal{A}(t)$ represents the asymmetry for molecules with different arrival times and hence different velocities.

As described in Ref.~\cite{Kara2012}, we collect 4096 consecutive shots into a ``block''. In each block, we switch a total of nine different parameters between two values, which allow us to calculate the eEDM, protect against systematic errors and noise, and optimise parameter values. To measure the eEDM, three parameters are crucial: $\mathtt{E}$, which sets the direction of the electric field; $\mathtt{B}$, which sets the magnetic field such that the interferometer phase is $\pm \pi/4$; and $\mathtt{\delta B}$, which adds or subtracts a small magnetic field $\Delta B$ to the one set by $\mathtt{B}$. For a switched parameter $\mathtt{X} = \{+1, -1\}$, we define the asymmetry correlated with the parameter as its ``channel'' $\{\mathtt{X}\}$. These channels retain the time-dependence of $\mathcal{A}(t)$. It can be shown that from a block, we obtain the eEDM as follows~\cite{Kara2012}:
\begin{equation}
\label{eq:edmFromAsymmetry}
    d_e(t) = \frac{g \mu_B \Delta B}{\eta E_\text{eff}} \frac{\{ \mathtt{E} \cdot \mathtt{B} \}}{\{ \mathtt{\delta B} \}}.
\end{equation}
The time-dependence of $d_e$ reminds us that for each block, we obtain measurements of $d_e$ over a range of arrival times of the molecules. 

\begin{figure}[tbp]
    \centering
    \includegraphics{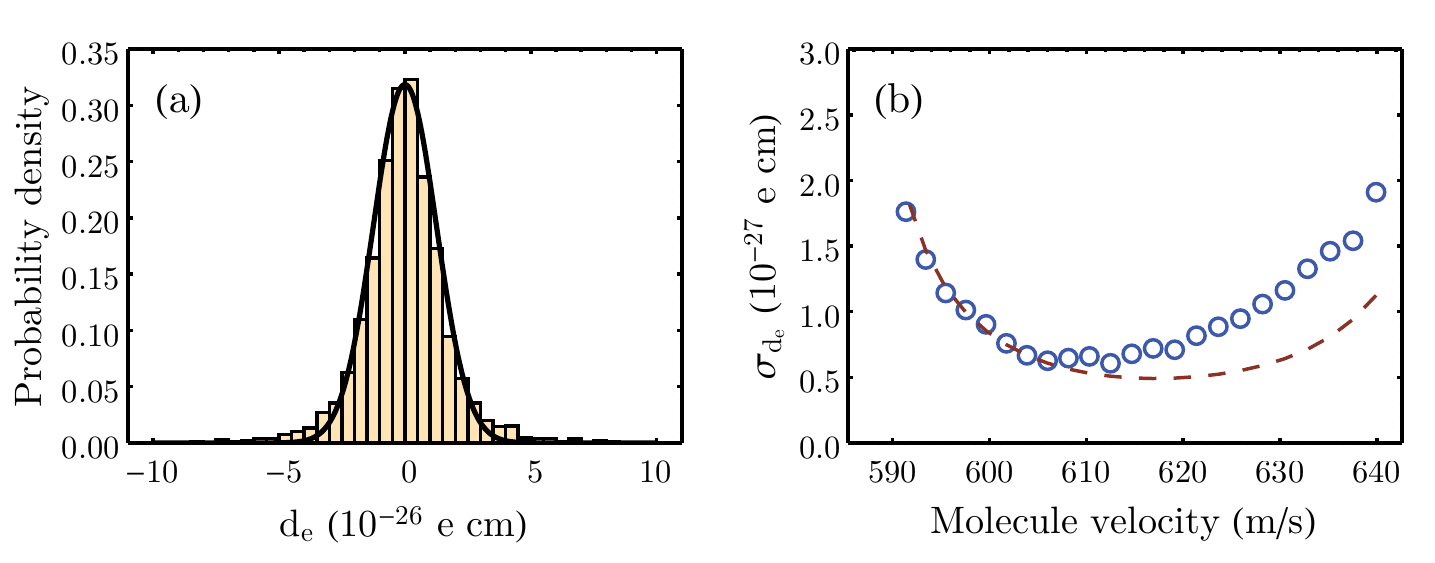}
    \caption{(a) Histogram of eEDM values from one day of measurement. (b) eEDM sensitivity as a function of molecule velocity (blue circles). The calculated shot-noise limit is shown by the red dashed line.}
    \label{fig:edmData}
\end{figure}

To demonstrate the sensitivity of the experiment, we took $N_\text{blocks} = 212$ blocks of data over a total of 22 hours, which we define to be one day of measurement time. We choose to analyse $d_e$ between molecule arrival times of $\SI{2675}{\mu s}$ and $\SI{2905}{\mu s}$, chosen to maximise the final sensitivity. These correspond to molecules with forward velocities between \SI{590}{m/s} and \SI{640}{m/s}. The resolution of data acquisition is $\SI{10}{\mu s}$, so we obtain 23 values of $d_e(t_i)$ per block, where $\{t_i\}$ are the arrival time bin centre values. A fixed, but unknown offset is added to all the values of $d_e$ to blind the analysis procedure. Figure~\ref{fig:edmData}(a) shows the distribution of eEDM values with their mean subtracted. The line is a normal distribution with zero mean and standard deviation equal to the average statistical uncertainty for each eEDM value, which is given by
\begin{equation}
    \label{eq:uncertaintyPerEDM}
    \sigma_{d_e, \text{single}} = \sqrt{N_\mathrm{v}}\ \sigma_{d_e},
\end{equation}
where $N_\mathrm{v} = 212 \times 23 = 4876$ is the total number of values of $d_e$ we include in the analysis, and $\sigma_{d_e}$ is the estimated standard error on determining the mean of $d_e$.

The standard error, $\sigma_{d_e}$, is obtained as follows. First, we divide the $N_\mathrm{v}$ values of $d_e$ into 23 equal datasets, grouped by their arrival time. We estimate the means of $d_e$ for each arrival time bin using the bootstrapped 10\% trimmed mean~\cite{Efron1986}. The trimmed mean is used because it is more robust to outliers, and bootstrapping is a random resampling technique used to easily obtain standard errors for parameter estimates even for non-normal distributions~\cite{Efron1986}. 

For each value of $t$, we have a set of $N_\text{blocks}$ values of $d_e$. From this set, we randomly draw $N_\text{blocks}$ values of $d_e$ to form a synthetic set, which we call a bootstrap set. We allow the same value to be drawn more than once. We then calculate the trimmed mean for the bootstrap set, which we call the bootstrap estimate of the trimmed mean. The procedure is repeated 5000 times, from which we obtain 5000 bootstrap estimates. The standard error on the trimmed mean is the standard deviation of these bootstrap estimates. We carry out the same procedure for each value of $t$, and finally take the weighted mean of the values of $d_e$ for the different $t$'s, where the weights are given by the inverse squares of the standard errors. The final uncertainty for $d_e$ is then given by the standard propagation of errors. We find that
\begin{equation}
\label{eq:edmsensitivity}
\sigma_{d_e} = \frac{1.8 \times 10^{-28}}{\sqrt{N_\text{days}}} \ \text{e cm},
\end{equation}
where $N_\text{days}$ is the number of days of measurement (where each day corresponds to 22 hours of measurement time). This is a factor of 20 better than the per-day sensitivity from the previous measurement \cite{Kara2012}.

It is useful to compare the experimental sensitivity to the shot-noise-limited sensitivity given in Eq.~\ref{eq:edmUncertaintyPiBy4}. Since we had subtracted away the background due to laser scatter from the apparatus, we now need to add in the noise in this background. Assuming that the latter is just shot noise, we get
\begin{equation}
    \label{eq:shotNoiseWithScatter}
    \xi_{d_e} = \frac{\hbar}{2 E_\text{eff} \tau \mathcal{C}} \sqrt{\frac{s_\text{T}+s_\text{scatter}}{s_\text{T}^2}},
\end{equation}
where $s_\text{scatter}$ is the total laser scatter measured in both detectors. To calculate the shot noise for a block, we sum over the contributions from all the shots, but retain the dependence on the arrival time $t_i$:
\begin{equation}
    \label{eq:snSensitivityPerBlock}
    \xi_{d_e, \text{block}}(t_i) = \frac{\hbar}{2 E_\text{eff} \tau \mathcal{C}(t_i)} \frac{1}{N_\text{shots}} \sqrt{\sum_{k=1}^{N_\text{shots}} \frac{s_{\text{T},k}(t_i) + s_{\text{scatter},k}}{\left( s_{\text{T},k}(t_i) \right) ^2}},
\end{equation}
where $s_{\text{T},k}$ and $s_{\text{scatter},k}$ are the total signal and laser scatter measured in the $k^\text{th}$ shot in the block, $N_\text{shots} = 4096$ is the number of shots in a block, and $\mathcal{C}(t_i)$ is the measured contrast as a function of arrival time from the block. For the dataset presented in this paper, this shot-noise-limited sensitivity is shown by the red dashed line in Fig.~\ref{fig:edmData}(b). In the same figure, the blue circles are the experimental uncertainties of $d_e$ for each molecule velocity, $v_i = L_2 / t_i$. This was calculated using the same bootstrapping method presented earlier. The increase in uncertainty for the molecules with higher speeds is currently under investigation. For the rest of the molecules, we are very close to the shot-noise limit. Carrying out a weighted sum of the shot-noise uncertainties, where the weights are the same as those used for the eEDM values, gives
\begin{equation}
    \label{eq:edmShotNoiseLimit}
    \xi_{d_e} = \frac{1.5 \times 10^{-28}}{\sqrt{N_\text{days}}} \ \text{e cm},
\end{equation}
indicating that the statistical sensitivity of the experiment is 1.2 times above the shot-noise limit.

\section{Conclusion and outlook}
\label{sec:conclusion}

The new state preparation and detection techniques presented in this paper, together with better collection optics in the detectors, have increased the number of photons detected by a factor of 400 since the previous eEDM measurement~\cite{Kara2012}. The state preparation technique used two separate lasers to address the $N=0$ and $N=2$ molecular populations, and a combination of microwave and rf fields to connect the $N=1$ population to $N=0$. By leaving out the $F=0$ component of the Q(0) laser, we are able to optically pump population into the desired ground state. After the interferometer region, the molecules in $F=0$ and $F=1$ were separately detected by connecting them to the $N=1$ levels and the detection efficiency was maximised by using the rotationally-closed P(1) transition. By modulating the polarisation of the detection lasers, counter-propagating the beams, and introducing a clean-up beam in the optical pumping region, we were able to maximise the contrast of the detector. These techniques are also important in the field of laser cooling molecules, where it is also necessary to scatter a large number of photons per molecule. Finally, we have demonstrated that the statistical sensitivity of the experiment is 1.2 times its shot-noise limit. Our current sensitivity is given by Eq.~\ref{eq:edmsensitivity} and is 20 times better than in our previous measurement. With this sensitivity, we would expect a statistical uncertainty at the \SI{e-29}{e.cm} level in a measurement consisting of 100 days of eEDM data. This is at a similar level of sensitivity to the current leading eEDM experiments~\cite{Cairncross2017,Andreev2018}, but with a different molecular species and experimental setup.

Further upgrades to the present apparatus are also possible. For example, we plan to introduce light tuned to the $\text{R}(1)$ transition just before each detection region, in order to remove population left behind in $N=1$ and therefore reduce the background. We also plan to use isotopically-pure $^{174}\text{Yb}$ for our molecular source, which will increase the number of useful molecules by a factor of three, since $^{174}\text{Yb}$ has a natural abundance of only 32\%. This will also remove the background signal due to transitions in other isotopologues of YbF, as discussed in Section~\ref{sec:contrast}. Furthermore, we have recently developed a buffer gas source of YbF and implemented transverse laser cooling~\cite{Lim2018}. This source provides a higher flux of molecules which have a lower forward velocity. Using a laser-cooled beam, much longer coherence times become feasible. The slower beam will also have a longer interaction time with the detection lasers, thereby giving a larger signal. Finally, the detection efficiency can be improved further by adding vibrational repump lasers to increase the number of photons scattered per molecule. The large expected increase in signal will also then make it favourable to switch from PMTs to silicon photomultipliers (SiPMs) which have a higher quantum efficiency, therefore increasing our overall detection efficiency. With all these proposed improvements, we expect to be able to probe the eEDM at the \SI{e-31}{e.cm} level.

\ack
This research has received support from the Royal Society and the Science and Technology Facilities Council (grants ST/N000242/1, ST/S000011/1). The research was also partly supported by the John Templeton Foundation (grant 61104), the Gordon and Betty Moore Foundation (grant 8864), and the Alfred P. Sloan Foundation (grant G-2019-12505).

\section*{References}
\bibliographystyle{iopart-num}
\bibliography{references}

\end{document}